\documentclass[amsmath, showpacs, twocolumn, superscriptaddress, prl]{revtex4-1}

\usepackage{txfonts}
\usepackage{bm}
\usepackage{graphicx,amsmath,mathrsfs}
\usepackage{amssymb}
\usepackage{float}
\usepackage[section]{placeins}
\usepackage{tabularx}
\usepackage{multirow}
\usepackage{color}
\usepackage{ifpdf}
\usepackage{graphicx, latexsym, CJK, bbm}
\usepackage{amsfonts}
\usepackage{booktabs}
\usepackage{enumerate}
\newcommand{\beq}{\begin{equation}}
\newcommand{\eeq}{\end{equation}}
\newcommand{\beqn}{\begin{eqnarray}}
\newcommand{\eeqn}{\end{eqnarray}}
\newcommand{\bea}{\begin{array}}
\newcommand{\eea}{\end{array}}
\newcommand{\bsub}{\begin{subequations}}
\newcommand{\esub}{\end{subequations}}
\newcommand{\bpm}{\begin{pmatrix}}
\newcommand{\epm}{\end{pmatrix}}

\newcommand{\scr}[1]{{\mathscr #1}}
\newcommand{\ff}[1]{\frac{1}{#1}}

\newcommand{\lrb}[1]{\Big(#1\Big)}
\newcommand{\lrs}[1]{\Big[#1\Big]}

\newcommand{\svec}[1]{{\mbox{\boldmath${ #1}$}}}
\newcommand{\ivec}{\vec}

\newcommand{\re}{\nonumber\\}

\newcommand{\sigs}{{\sigma\text{-S} }}
\newcommand{\omev}{{\omega\text{-V} }}
\newcommand{\rhov}{{\rho  \text{-V} }}
\newcommand{\rhot}{{\rho  \text{-T} }}

\newcommand{\pipv}{{\pi   \text{-PV}}}

\usepackage{ulem}
\newcommand{\delete}{\bgroup\markoverwith{\textcolor{red}{\rule[0.5ex]{2pt}{1pt}}}\ULon}

\allowdisplaybreaks

\begin{document}

\title{ The Nuclear Tensor Interaction in Covariant Energy Density Functional }

\author{Li Juan Jiang}
\author{Shen Yang}
\author{Bao Yuan Sun}
\author{Wen Hui Long}\email{longwh@lzu.edu.cn}
\author{Huai Qiang Gu}
\affiliation{School of Nuclear Science and Technology, Lanzhou University,730000 Lanzhou, China}

\begin{abstract}
  The origin of the nuclear tensor  interaction in the covariant energy density functional (EDF) is presented in this work, associated with the Fock diagrams of Lorentz scalar and vector couplings. With this newly obtained relativistic formalism of the nuclear tensor interaction, more distinct tensor effects are found in the Fock diagrams of the Lorentz scalar and vector couplings, as compared to the Lorentz pseudo-vector and tensor channels. A unified and self-consistent treatment on both  the nuclear tensor and spin-orbit interactions, which dominate the spin-dependent features of the nuclear force, is then achieved by the relativistic models. Moreover, careful analysis on the tensor strengths indicates the reliability of the nuclear tensor interaction in the covariant EDF for exploring the nuclear structure, excitation and decay modes.
\end{abstract}


\maketitle

Since the birth of nuclear physics, the nuclear force that binds protons and neutrons into an atomic nucleus is the most significant issue of the field. The earliest attempt in understanding the nature of the nuclear force was made by Yukawa with the meson exchange picture \cite{Yukawa:1935}.  To a large extent, the nuclear force can be understood in terms of the exchanges of virtual mesons, which is the microscopic foundation of modern nuclear theories, such as the covariant density functional (CDF) theory \cite{Serot:1986}. At a very early stage, the nuclear force was recognized to contain not only central components but also the non-central ones, i.e., the nuclear tensor force that plays an essential role in binding the light nuclei \cite{Gerjuoy1942, Feshbach1949, Schrenk1967, Pudliner1997}. Specifically, the electric quadrupole moment of the deuteron provides the most striking evidence of he nuclear tensor interaction \cite{Fayache1997}.

As an important ingredient of the nuclear force, the  nuclear tensor interaction is characterized by its spin dependent feature \cite{Otsuka2005}. In the recent years, substantial impacts due to the nature of the tensor force were recognized in the extensions of the nuclear chart from traditional stable nuclei to exotic ones \cite{Otsuka2005, Colo2007, Lesinski2007, Zuo2008, Otsuka2010, Anguiano:2012PRC054302}. Moreover, impressive progresses associated with the nuclear tensor force were also achieved in describing the nuclear excitations \cite{Bai2009PLB, Bai2010PRL, Cao2009PRC, Anguiano:2011PRC064306, G.CO:2012PRC034323} and decay modes \cite{Minato2013}. For instance, within the Skyrme Hartree-Fock (SHF) plus random phase approximation (RPA) scheme, it was found that the tensor force components play a crucial role in understanding the Gamow-Teller (GT) transition \cite{Bai2009PLB}, charge exchange spin-dipole (SD) excitations \cite{Bai2010PRL}, the non-charge exchange multipole responses \cite{Cao2009PRC} and the $\beta$-decay of magic and semi-magic nuclei \cite{Minato2013}. Besides, the tensor force was also found to have substantial effects in determining the density-dependent behavior of the symmetry energy \cite{Baoan2010, Vidana2011} {that is} the key quantity in understanding the nuclear equation of state and relevant astrophysical processes \cite{Baoan2008, Lattimer2004}.

Usually, the nuclear tensor interaction is identified by the following form,\vspace{-0.5em}
  \begin{equation}\label{Tensor_Wigner}
     S_{12}= 3(\svec\sigma_1\cdot\svec q) (\svec\sigma_2\cdot\svec q) - \svec\sigma_1\cdot\svec\sigma_2\svec q^2,
  \end{equation}
where $S_{12}$ is a rank-2 tensor operator well defined in the non-relativistic quantum mechanics, with the momentum transfer $\svec q=\svec p_1-\svec p_2$. While there still remain some unresolved problems, such as the origin of the nuclear tensor force and its coupling strength. For the later there exists an evident model dependence with respect to the widely used energy functionals such as the Skyrme forces \cite{Sagawa2014Colo}. Within the CDF scheme, which provides a self-consistent treatment on the spin-orbit coupling, several attempts {were also} made to explore the tensor effects, e.g., in terms of $\omega$-tensor couplings \cite{Mao2003PRC}. However, these are Lorentz tensors and they give pure central type contributions in the limit of Hartree approach. Under the meson exchange picture, the nuclear tensor force was recognized to originate from the exchanges of $\pi$ and $\rho$ (mainly tensor $\rho$) mesons \cite{Otsuka2005, Long2008}. However, only when the Fock terms of meson-nucleon couplings are included explicitly, the $\pi$ and $\rho$-tensor couplings can be  efficiently taken into account, for instance, by the density dependent relativistic Hartree-Fock (DDRHF) theory \cite{Long640, Long2007, Long2010},  from which distinct tensor effects are revealed in nuclear structure properties \cite{Long2007, Long2008, Wang2013}. {Even though, the Fock terms of the Lorentz tensor couplings, e.g., the $\pi$ pseudo-vector and $\rho$ tensor couplings, are still mixtures of the central and tensor force components \cite{Long2008}.}

Furthermore, a fully self-consistent charge-exchange relativistic RPA based on DDRHF, namely the DDRHF+RPA model, has been established to describe the spin-isospin resonances like GT and SD ones, from which is well demonstrated the crucial role played by the exchange (Fock) diagrams of the isoscalar $\sigma$ and $\omega$ couplings \cite{Liang2008, Liang2012}. Notice that these excitation modes were interpreted successfully by the Skyrme+Tensor models as well \cite{Bai2009PLB, Bai2010PRL}, in which the tensor force was found to play a key role. As an indirect evidence, such consensus indicates that the tensor force components may exist in the Fock diagrams of meson-nucleon couplings,  not only the isovector ones ($\pi$ and $\rho$) but also the isoscalar ones ($\sigma$ and $\omega$).

\begin{figure}[htbp]
\includegraphics[width = 0.48\textwidth]{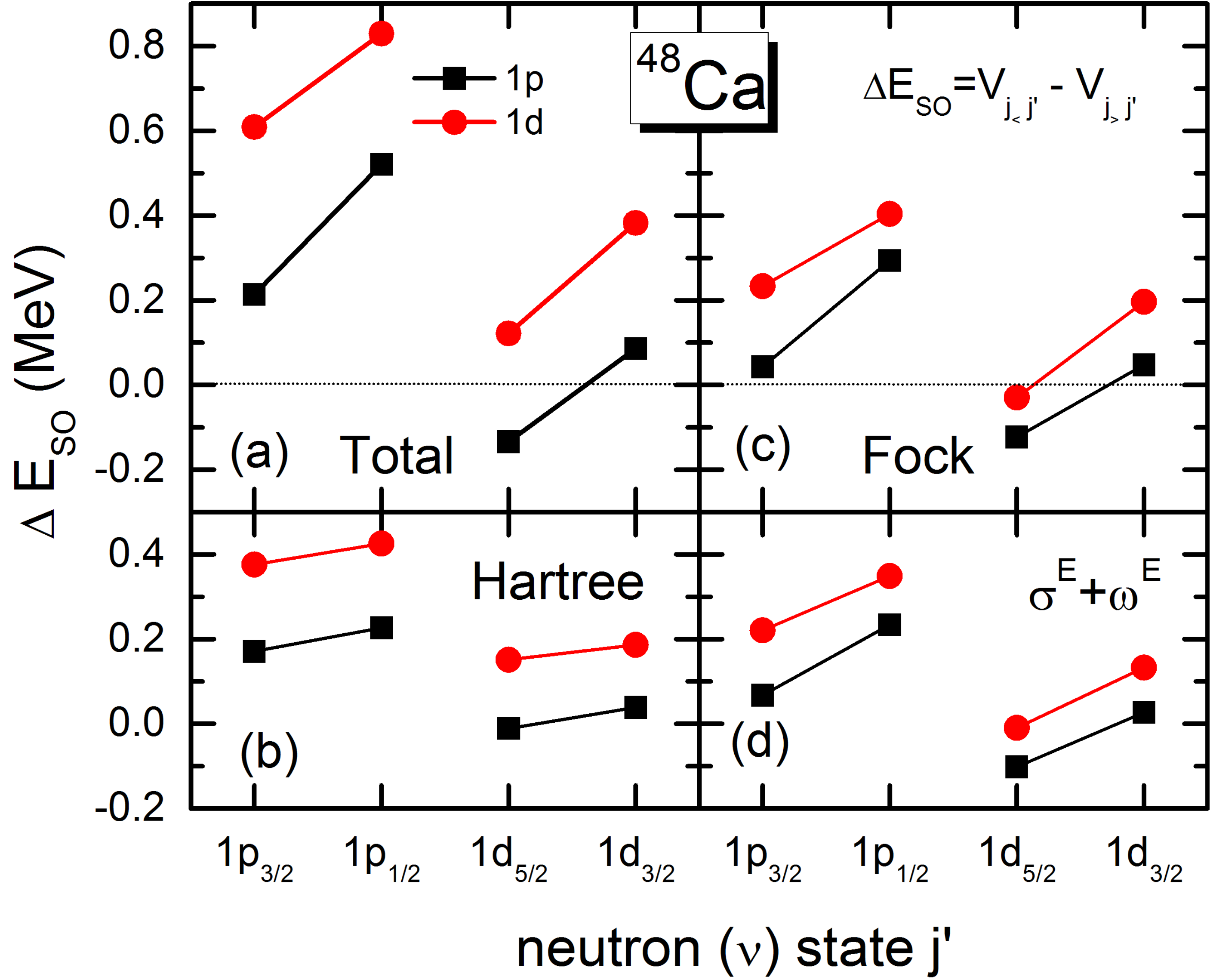}
\caption{(Color Online) Contributions to the spin-orbit splittings $\Delta E_{\rm SO} = V_{j_<j'} - V_{j_>j'}$ (MeV) of the nodeless neutron ($\nu$) orbits ($\nu1p$ and $\nu1d$) from the couplings with the neutron on the nodeless states ($j'$) in $^{48}$Ca. In the plots (a-d) are shown the contributions of the total, Hartree terms, Fock terms, and the Fock terms of $\sigma$- and $\omega$-couplings (namely $\sigma^E + \omega^E$), respectively. The result are extracted from the calculations of DDRHF with PKA1.}
\label{Pb208SO}
\end{figure}

In fact, when the Fock diagrams are included, the {nuclear force mediated by meson exchanges} is found to contain the characteristic spin-dependence of a tensor force. Associated with the nature of tensor force \cite{Otsuka2005}, the spin-orbit (SO) splitting will be essentially changed by the tensor couplings [see Eq. (\ref{Tensor_Wigner})], thus providing a direct test for the existence of nuclear tensor interaction.  To simplify the notation, we take the SO splittings of neutron ($\nu$) $p$ and $d$ orbits of $^{48}$Ca as the test examples. Figure \ref{Pb208SO} (a-d) shows the contributions to the SO splittings ($\Delta E_{\rm SO} = V_{j_< j'} - V_{j_>j'}$) respectively from the neutron-neutron interactions of the total, the Hartree and Fock terms, and the Fock terms of the isoscalar $\sigma$- and $\omega$-meson couplings (denoted by $\sigma^E +\omega^E$). It is seen that the total $\Delta E_{\rm SO}$ are essentially changed from $j' = l'+1/2$ to $l'-1/2$, which indicates that the neutron-neutron interactions are distinctly spin-dependent. In addition, such characteristic behaviors are dominated by the Fock diagrams, particularly the isoscalar contributions $\sigma^E +\omega^E$. This provides a concrete evidence for the existence of the tensor force components in the Fock diagram of meson-nucleon couplings, particularly in the isoscalar channels. On the other hand, it is confirmed that the tensor terms (\ref{Tensor_Wigner}) are also found in the non-relativistic reduction of the Fock terms of isoscalar meson-nucleon couplings, similar as the isovector ones \cite{Bouyssy1987}. Therefore, the Fock diagrams can be considered as the mixture of central and tensor force contributions, not only for the Lorentz tensor --- $\pi$ pseudo-vector (PV) and $\rho$ tensor (T) couplings \cite{Bouyssy1987, Long2007, Long2008} but also for the Lorentz $\sigma$ scalar (S) and $\omega$ vector (V) ones, the new origin of nuclear tensor force.

Notice that the spin operator $\hat{\svec S} = \ff2\svec\sigma$ in $S_{12}$ (\ref{Tensor_Wigner}) can be identified relativistically  as $\hat{\svec S} = \ff2\svec\Sigma = -\ff2\gamma_0\gamma_5\svec\gamma$, and $\gamma_5\svec\gamma$ is the Dirac index of $\pi$-PV coupling. Inspired by the extraction of tensor contributions in the one-pion exchange potential \cite{Long2008} and the non-relativistic reductions of the Fock terms, we present the following relativistic formalism to extract the tensor force components hiding in the energy density functionals (EDF) of the $\pi$-PV, $\sigma$-scalar (S), $\omega$-vector (V) and $\rho$-tensor (T) couplings, 
\begin{align}
\scr H_\pipv^T = & -\ff2\lrs{ \frac{f_\pi}{m_\pi} \bar\psi  \gamma_0 \Sigma_\mu\ivec\tau \psi }_1\cdot\lrs{\frac{f_\pi}{m_\pi} \bar\psi\gamma_0 \Sigma_\nu \ivec\tau \psi}_2 D_\pipv^{T,\ \mu\nu}(1,2),\label{Wigner-tensor-pi-R}\\
\scr H_\sigs^T = & -\ff2\cdot\ff2\lrs{\frac{g_\sigma}{m_\sigma}\bar\psi \gamma_0\Sigma_\mu\psi }_1\lrs{\frac{g_\sigma}{m_\sigma}\bar\psi \gamma_0\Sigma_\nu \psi }_2 D_\sigs^{T,\ \mu\nu}(1,2),\label{Wigner-tensor-sig-R}\\
\scr H_\omev^T = & +\ff2\cdot\ff2\lrs{\frac{g_\omega}{m_\omega}\bar\psi \gamma_\lambda\gamma_0\Sigma_\mu\psi }_1 \lrs{\frac{g_\omega}{m_\omega}\bar\psi \gamma_\delta \gamma_0\Sigma_\nu \psi }_2 D_\omev^{T,\ \mu\nu\lambda\delta}(1,2),\label{Wigner-tensor-ome-R}\\
\scr H_\rhot^T = &+ \ff2 \lrs{\frac{f_\rho}{2M}\bar\psi \sigma_{\lambda\mu}\ivec\tau\psi }_1\cdot \lrs{\frac{f_\rho}{2M}\bar\psi \sigma_{\delta\nu}\ivec\tau\psi }_2 D_\rhot^{T,\ \mu\nu\lambda\delta} (1,2),\label{Wigner-tensor-rhot-R}
\end{align}
where the additional factor 1/2 in $\scr H_\sigs^T$ and $\scr H_\omev^T$ originates from the non-relativistic reduction of the relevant Fock terms, $\Sigma^\mu = \big(\gamma^5, \svec\Sigma\big)$, $M$ is the nucleon mass, and $\ivec\tau$ denotes the isospin operator of the nucleon ($\psi$). The propagator terms $D^T$ read as,
\begin{align}
D_\phi^{T,\ \mu\nu}(1,2) = &\lrs{\partial^\mu(1)\partial^\nu(2) - \ff3 g^{\mu\nu} m_\phi^2} D_\phi (1,2)\re
&\hspace{4.8em} + \ff3 g^{\mu\nu} \delta(x_1-x_2), \\
D_{\phi'}^{T,\ \mu\nu\lambda\delta}(1,2) = & \partial^\mu(1)\partial^\nu(2) g^{\lambda\delta}  D_{\phi'}(1,2) \re
&- \ff3 \lrb{g^{\mu\nu} g^{\lambda\delta} - \ff3 g^{\mu\lambda} g^{\nu\delta} } m_{\phi'}^2 D_{\phi'}(1,2)\re
&+ \ff3 \lrb{g^{\mu\nu} g^{\lambda\delta} - \ff3 g^{\mu\lambda} g^{\nu\delta} } \delta(x_1-x_2),\label{propagator-V}
\end{align}
where $\phi$ stands for the $\sigma$-S and $\pi$-PV couplings, and $\phi'$ represents the $\omega$-V and $\rho$-T channels. For the $\rho$-V coupling, $\scr H_\rhov^T$, a corresponding formalism can be obtained simply by replacing $m_\omega$ ($g_\omega$) in eqs. (\ref{Wigner-tensor-ome-R}) and (\ref{propagator-V}) by $m_\rho$ ($g_\rho$) and inserting the isospin operator $\ivec\tau$ in the interacting index. In keeping with the theory itself, the $\mu,\nu=0$ components of the propagator terms will be omitted in practice, which amounts to neglecting the retardation effects.  Transferring to the momentum space, the interaction index together with the propagator term  in $\scr H_\phi^T$ ($\phi=\sigs$ and $\pipv$) can be expressed as,
\begin{align}
  V_\phi^T(\svec q) = & \ff3\frac{3\big(\gamma_0\svec\Sigma_1\cdot\svec q\big)\big(\gamma_0\svec\Sigma_2\cdot\svec q\big) -  \big(\gamma_0\svec\Sigma_1\big)\cdot\big(\gamma_0\svec\Sigma_2\big)\svec q^2}{m_\phi^2+\svec q^2},
\end{align}
and the numerator term in the right-hand side is exactly a rank-2 irreducible tensor operator similar as $S_{12}$ [see Eq. (\ref{Tensor_Wigner})]. For $\phi'=\omev$, $\rhot$ and $\rhov$, one may obtain the irreducible tensor operators with higher ranks.

\begin{figure}[htbp]
\includegraphics[width = 0.48\textwidth]{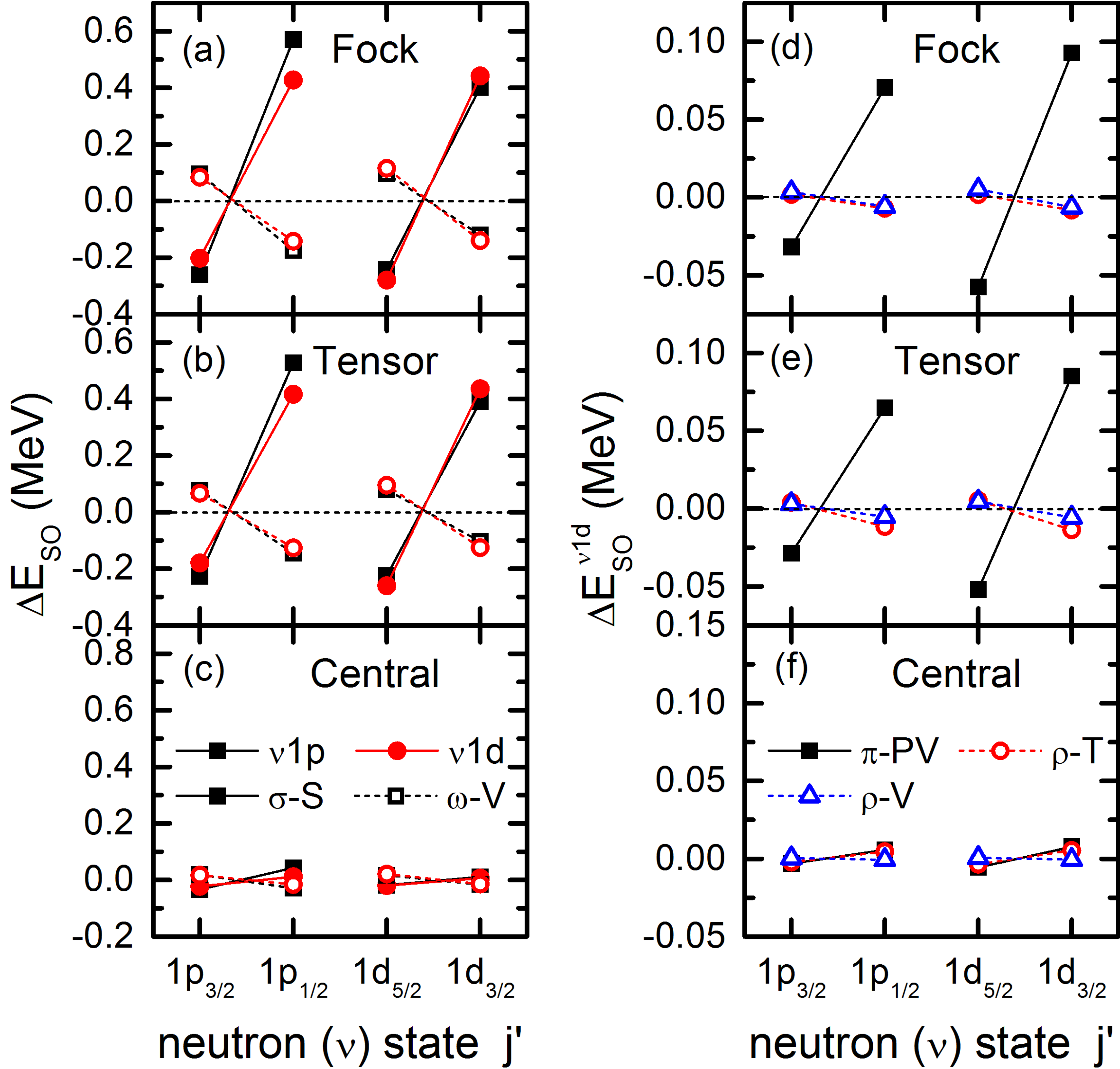}
\caption{(Color Online) Contributions to the spin-orbit splittings $\Delta E_{\text{SO}} = V_{j_<j'} - V_{j_>j'}$ (MeV) from the Fock diagrams [plots (a, d)], and their tensor [plots (b, e)] and central [plots (c, f)] parts. The results are extracted from the calculations of DDRHF functional PKA1 \cite{Long2007} by taking the nodeless neutron ($\nu$) orbits in $^{48}$Ca as examples and the spin partner states $j_>$ and $j_<$ share the same radial wave function. In plots (a-c) the filled (open) symbols denote the contributions from $\sigma$-S ($\omega$-V) couplings. In plots (d-f) are only shown the results of the nodeless neutron orbit $\nu1d$ for $\pipv$, $\rhov$ and $\rhot$ couplings.
}
\label{Pb208SOTC}
\end{figure}

To test the validity of the proposed formalism [eqs. (\ref{Wigner-tensor-pi-R}-\ref{Wigner-tensor-rhot-R})] as the relativistic representation of the nuclear tensor interaction in the covariant EDF, Fig. \ref{Pb208SOTC} shows the relevant contributions to the SO splittings $\Delta E_{\text{SO}}$ of nodeless $\nu1p$ and $\nu1d$ orbits of $^{48}$Ca, namely the total Fock terms [plot (a,d)], the tensor [plot (b, e)] and remaining central parts [plot (c, f)]. The calculations are performed with the DDRHF functional PKA1 which presents more complete RHF scheme of meson-nucleon couplings \cite{Long2007} than the PKO series \cite{Long640, Long2008}. In order to clearly identify the tensor effects, the same radial wave functions are used for the spin partner states $j_>$ and $j_<$ in calculating the interacting matrix elements $V_{j_\gtrless j'}$. With the restriction, it is found that the contributions to $\Delta E_{\text{SO}}$ from the Fock terms act like the nuclear tensor force  [see Fig. \ref{Pb208SOTC}(a, d)], and the tensor feature --- the spin dependence can be extracted and quantified almost completely by the relativistic formalism [see Fig. \ref{Pb208SOTC}(b, e)]. Not only the contributions to the SO splittings, the interacting matrix elements determined by the relativistic formalism (\ref{Wigner-tensor-pi-R}-\ref{Wigner-tensor-rhot-R}), i.e., $V_{j_>j_>'}^T$ ( or $V_{j_<j_<'}^T$) are also found opposite to those $V_{j_>j'_<}^T$ (or $V_{j_<j'_>}^T$), consistent with the nature of tensor force \cite{Otsuka2005}. Besides, the tensor effects contributed by the Fock diagrams of $\sigs$ ($\pipv$) and $\omev$ ($\rhov$ and $\rhot$) couplings are opposite and counteracted by each another, similarly to the cancellation between strong $\sigma$-attraction and $\omega$-repulsion. Compared to the isovector channels ($\pi$-PV, $\rho$-V and $\rho$-T), more distinct tensor effects, with almost one order of magnitude larger, are brought about by the isoscalar ones ($\sigma$-S and $\omega$-V), consistent with the results shown in Fig. \ref{Pb208SO}(c, d).

\begin{table}[floatfix]
\caption{Interaction matrix elements $V_{j_\gtrless j'}^T$ of the tensor force components in the Fock diagrams of $\sigma$-S and $\omega$-V couplings with the limits that the spin partner states $j_<$ and $j_>$ share the same radial wave functions and the contributions of the small components of Dirac spinors are omitted. The results are extracted from the calculations of DDRHF with PKA1 for the neutron ($\nu$) orbits of $^{48}$Ca. }\label{Pb208IME_T}
\begin{tabular}{c|rrrr|rrrr}  \hline\hline
\multirow{2}{*}{$V_{j_\gtrless j'}^T$} &\multicolumn{4}{c|}{$\sigma$-S ($10^{-1}$MeV)} &    \multicolumn{4}{c}{$\omega$-V ($10^{-1}$MeV)}\\[0.25em]
              &$\nu1p_{1/2}$& $\nu1d_{5/2}$&$\nu1d_{3/2}$& $\nu1f_{7/2}$&$\nu1p_{1/2}$& $\nu1d_{5/2}$&$\nu1d_{3/2}$& $\nu1f_{7/2}$\\   \hline
$\nu1p_{ 3/2}$&      $-$1.72&          0.80&      $-$1.24&          0.56&         0.54&       $-$0.26&         0.41&       $-$0.19\\
$\nu1p_{ 1/2}$&         3.43&       $-$1.60&         2.48&       $-$1.11&      $-$1.08&          0.53&      $-$0.82&          0.39\\   \hline
$\nu1d_{ 5/2}$&      $-$1.62&          1.13&      $-$1.66&          1.02&         0.54&       $-$0.38&         0.56&       $-$0.36\\
$\nu1d_{ 3/2}$&         2.44&       $-$1.69&         2.50&       $-$1.53&      $-$0.81&          0.57&      $-$0.85&          0.54\\
\hline\hline
\end{tabular}
\end{table}

\begin{figure}[htbp]
\includegraphics[width = 0.45\textwidth]{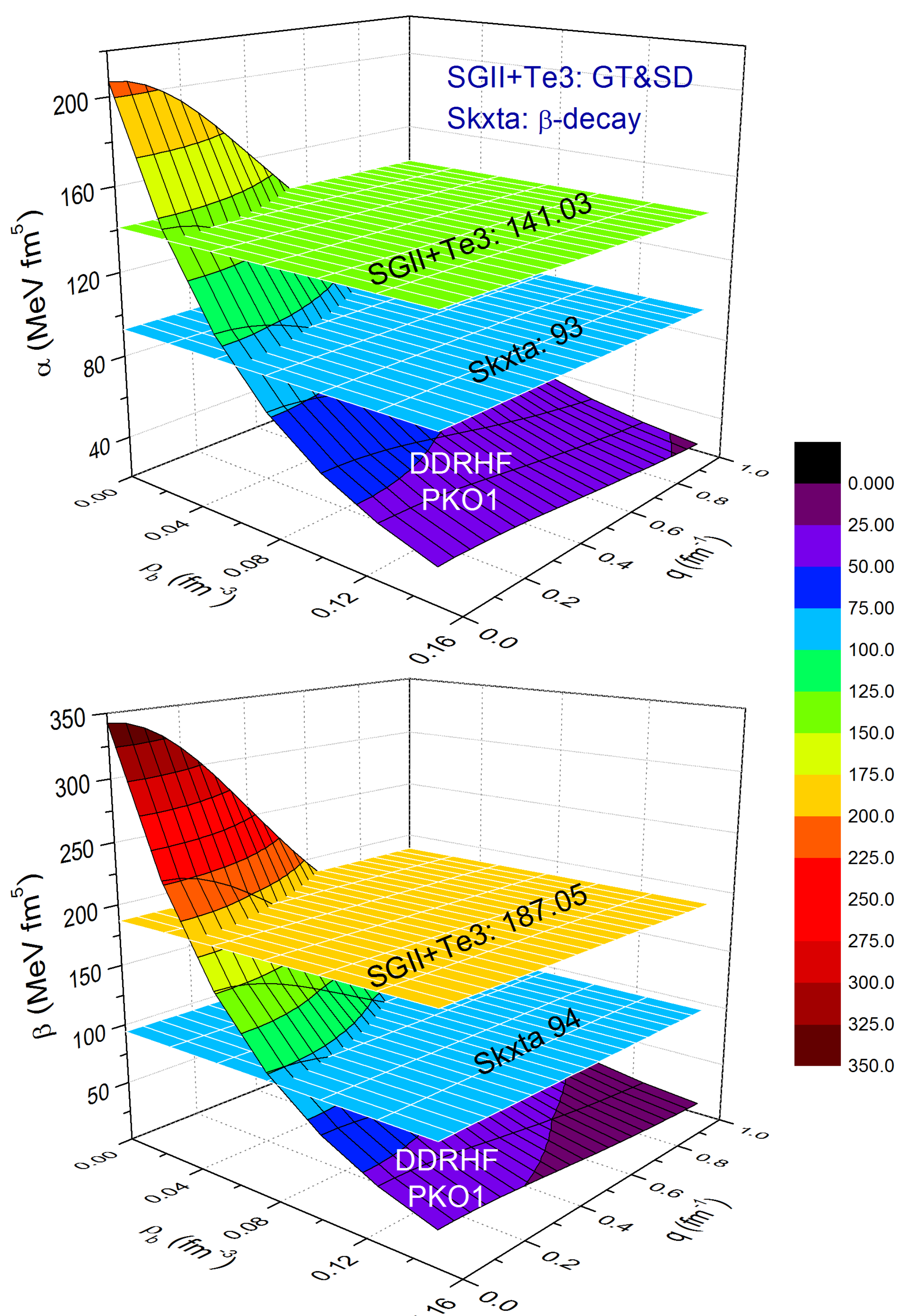}
\caption{(Color Online) Tensor strength factors $\alpha$ and $\beta$ (MeV$\cdot$fm$^{5}$) with respect to nucleon density $\rho_b$ (fm$^{-3}$) and momentum transfer $q$. The results are extracted from the non-relativistic reduction of relativistic representation for the tensor force components in DDRHF functional PKO1, as compared to the ones determined by the Skyrme forces SGII+Te3 \cite{Bai2011PRC} and Skxta \cite{Brown2006}.}
\label{O1-ABT}
\end{figure}

As {a test}, the tensor sum rule $(2j_>+1) V_{j_>j'}^T + (2j_<+1) V_{j_<j'}^T=0$ \cite{Otsuka2005} is verified with the relativistic formalism [Eqs. (\ref{Wigner-tensor-pi-R}-\ref{Wigner-tensor-rhot-R})]. Taking the neutron ($\nu$) orbits of $^{48}$Ca as examples, Table \ref{Pb208IME_T} shows the interaction matrix elements $V_{j_\gtrless j'}^T$ calculated by the relativistic formalism (\ref{Wigner-tensor-sig-R}-\ref{Wigner-tensor-ome-R}) of $\sigma$-S and $\omega$-V channels and the calculations are performed with the limit that the spin partner states $j_>$ and $j_<$  (the first column in Table \ref{Pb208IME_T}) share the same radial wave function \cite{Otsuka2005} and the small components of Dirac spinors are omitted. It is found that the  tensor sum rule is exactly fulfilled under this limit. Similar tests are also performed for the relativistic formalism (\ref{Wigner-tensor-pi-R}, \ref{Wigner-tensor-rhot-R}) of $\pipv$ and $\rhot$ channels as well as the $\rhov$ one, and the tensor sum rules are obeyed in this limit.

On the other hand, it should be noticed that a nuclear tensor interaction emerges simultaneously with the presence of Fock diagrams in the covariant EDF and the relevant tensor effects can be extracted completely by the proposed relativistic formalism [Eqs. (\ref{Wigner-tensor-pi-R}-\ref{Wigner-tensor-rhot-R})] without introducing any additional free parameters. From this point of view, the advantage of full relativistic Hartree-Fock (RHF) scheme based on meson exchange diagram of nuclear force, is then well demonstrated. Namely, the unified and self-consistent treatment of both tensor and SO interactions can be achieved by the RHF scheme, respectively due to the Fock diagrams and Lorentz covariant structure of the theory itself. Moreover,  with the relativistic representation of the tensor force components [i.e., Eqs. (\ref{Wigner-tensor-pi-R}-\ref{Wigner-tensor-rhot-R})], direct constraints from the tensor-related observables are then feasible to optimize the relativistic EDF, which may also promote our understanding on the nature of nuclear force.

Not only {on} nuclear ground states \cite{Long2008, Colo2007}, but also in nuclear excitations \cite{Liang2008, Liang2012, Bai2009PLB, Bai2010PRL} and $\beta$-decay \cite{Minato2013, Niu2013} there is a common understanding of the non-relativistic and relativistic models, for instance the SHF and RHF models. Both indeed share the success due to the presence of the tensor force component which is added to the Skyrme EDF, or naturally involved in the RHF one. For the non-relativistic SHF models, the tensor contributions to the SO potential may originate from the added tensor terms and the exchange part of the central Skyrme interaction and the tensor strength factors are determined as $\alpha = \alpha_T + \alpha_C$ and $\beta = \beta_T + \beta_C$ \cite{Colo2007}. From the non-relativistic reduction of the relativistic formalism (\ref{Wigner-tensor-pi-R}-\ref{Wigner-tensor-rhot-R}), these strength factors can be determined approximately as,
\begin{subequations}\label{alpha_beta}
\begin{align}
\alpha=& \frac{5}{12}\Big\{\ff4 \frac{g_\sigma^2}{m_\sigma^2} \ff{m_\sigma^2 + \svec q^2} -\ff4 \frac{g_\omega^2}{m_\omega^2} \ff{m_\omega^2+\svec q^2}+ \ff2 \frac{f_\pi^2}{m_\pi^2} \ff{m_\pi^2 + \svec q^2}  \re
& \hspace{5em} -\lrs{\ff4 \frac{g_\rho^2}{m_\rho^2} - \ff2\frac{f_\rho^2}{4M^2}} \ff{m_\rho^2+\svec q^2}  \Big\}, \\
\beta=&\frac{5}{6}\lrs{ \ff2 \frac{f_\pi^2}{m_\pi^2} \ff{m_\pi^2 + \svec q^2} -\lrb{\ff4 \frac{g_\rho^2}{m_\rho^2} - \ff2\frac{f_\rho^2}{4M^2}}\ff{m_\rho^2+\svec q^2} },
\end{align}
\end{subequations}
which depend on momentum transfer $\svec q$ due to the Yukawa propagators of meson exchanges and the baryon density $\rho_b$ if the meson-nucleon couplings ($g_\sigma$, $g_\omega$, $g_\rho$, $f_\pi$ and $f_\rho$) are density-dependent. In the above expressions, the contributions of higher order terms are eliminated, e.g., the space  components of $\scr H_\omev^T$ [see Eq. (\ref{Wigner-tensor-ome-R})] and $\scr H_\rhov^T$ are of the order of $1/M^2$, as well as the time component of $\scr H_\rhot^T$ [see Eq. (\ref{Wigner-tensor-rhot-R})].

Notice that the Skyrme forces SGII+Te3 \cite{Bai2011PRC} and Skxta \cite{Brown2006} are very successful respectively in describing nuclear excitations \cite{Bai2011PRC, Bai2011O16} and $\beta$-decay \cite{Minato2013}, whereas the DDRHF functional PKO1 \cite{Long640} succeeds in both cases \cite{Liang2008, Liang2012, Niu2013}. Figure \ref{O1-ABT} shows the tensor strength factors $\alpha$ and $\beta$ with respect to baryon density $\rho_b$ and momentum transfer $q$ determined by PKO1, in comparison with SGII+Te3 and Skxta.  For SGII+Te3, the similarities with the tensor strengths determined by PKO1 are found in lower density region with narrower range of momentum transfer $q$, as compared to Skxta. In fact, based on an existing Skyrme functional like SGII, distinct uncertainty still remains in determining the tensor strengths even with the constraint of the spin-isospin resonances \cite{Bai2011PRC}. In this work, the strength factors $\alpha$ and $\beta$ are extracted directly from the DDRHF functionals [see Eq. (\ref{alpha_beta})] which were developed by the fittings of the nuclear binding energies, radii, etc. Meanwhile, due to the fact that the tensor force components in the relativistic EDF are the innate parts of the Fock diagrams, it is then expected that the tensor strengths can be also constrained properly by the parametrization of the DDRHF functionals. In practice, such expectation is illustrated by the fact that the DDRHF+RPA model with the existing DDRHF functionals provides a full self-consistent covariant description of the spin-isospin resonances \cite{Liang2008, Liang2012}, being successful in describing the $\beta$-decay as well \cite{Niu2013}. In contrast to the zero-range tensor terms added to the Skyrme EDF, the tensor components involved automatically by the Fock diagrams in the covariant EDF may have some advantage in the extensive applications, due to the fact that important correlations are taken into account simultaneously, for instance, the nuclear in-medium effects evaluated by the density dependence of the tensor couplings and the finite-range features carried by the Yukawa-type propagators.

In summary, the relativistic representation of the nuclear tensor interaction in the covariant energy density functional (EDF) is proposed with the new origin associated with the Fock diagrams of Lorentz scalar ($\sigma$ and $\delta$) and vector ($\omega$ and $\rho$) couplings. The proposed relativistic formalism, which are utilized to quantify the tensor feature carried by the Fock diagrams of meson-nucleon couplings, are confirmed to be identical with the nature of tensor force, in terms of the spin-orbit interactions as well as the tensor sum rule.  Specifically more distinct tensor effects are found in the isoscalar than the isovector channels, which may interpret the success achieved by the DDRHF+RPA scheme in describing nuclear excitation modes. Due to the self-consistence on involving the nuclear tensor interaction into the covariant EDF, unified and self-consistent treatment on both tensor and spin-orbit interactions can be achieved by the relativistic models with the presence of Fock diagrams, which is of special meaning in exploring the limits of existence of nuclear systems. Moreover, with the careful analysis on the tensor strengths ($\alpha$ and $\beta$) determined by the relativistic model (DDRHF-PKO1) and non-relativistic ones (e.g., SGII+Te3 and Skxta) and the common successes achieved by both models, it well demonstrates the reliability of the relativistic representation of the nuclear tensor force in describing nuclear structure, excitation and decay modes. 

We would like to thank Prof. N. Van Giai, Prof. J. Meng and Prof. G. Col$\grave{\rm o}$ for their enlightening discussions and fruitful helps. This work is partly supported by the National Natural Science Foundation of China under Grant Nos. 11375076 and 11205075, and the Specialized Research Fund for the Doctoral Program of Higher Education under Grant Nos. 20130211110005 and 20120211120002.
%


%

\end{document}